# The Effect of Geometry Parameters and Flow Characteristics on Erosion and Sedimentation in Channel's Junction using Finite Volume Method

Mohammadamin Torabi[1], Amirmasoud Hamedi[2], Ebrahim Alamatian[3] and Hamidreza Zahabi[4]
[1]Ph.D. Candidate, Department of Civil and Environmental Engineering, Idaho State University, Pocatello, ID, USA
[2]Ph.D, Civil and Environmental Engineering Department, Florida International University, FLORIDA
[3]Assistant Professor, Khavaran Institute of Higher Education, Mashhad, IRAN
[4]Ph.D, Department of Civil Engineering, Institute Superior Tecnico, Lisbon, PORTUGAL

[4]Corresponding Author: hamidreza.zahabi@tecnico.ulisboa.pt

**ABSTRACT**
One of the most critical problems in the river engineering field is scouring, sedimentation and morphology of a river bed. In this paper, a finite volume method FORTRAN code is provided and used. The code is able to model the sedimentation. The flow and sediment were modeled at the interception of the two channels. It is applied an experimental model to evaluate the results. Regarding the numerical model, the effects of geometry parameters such as proportion of secondary channel to main channel width and intersection angle and also hydraulic conditionals like secondary to main channel discharge ratio and inlet flow Froude number were studied on bed topographical and flow pattern. The numerical results show that the maximum height of bed increased to 32 percent as the discharge ratio reaches to 51 percent, on average. It is observed that the maximum height of sedimentation decreases by declining in main channel to secondary channel Froude number ratio. On the assessment of the channel width, velocity and final bed height variations have changed by given trend, in all the ratios. Also, increasing in the intersection angle accompanied by decreasing in flow velocity variations along the channel. The pattern of velocity and topographical bed variations are also constant in any studied angles.

## I. INTRODUCTION

Flow and depth variations are influential factors on bed sedimentation and erosion in separation zone placed on channels intersection. At the junction of two streams, Best's studies showed that the location of rivers intersection divided to six different zones (Fig. 1). The zones are including stagnation zone, flow deflection zone, flow separation zone, maximum velocity zone, flow recovery zone and shear layers zone.

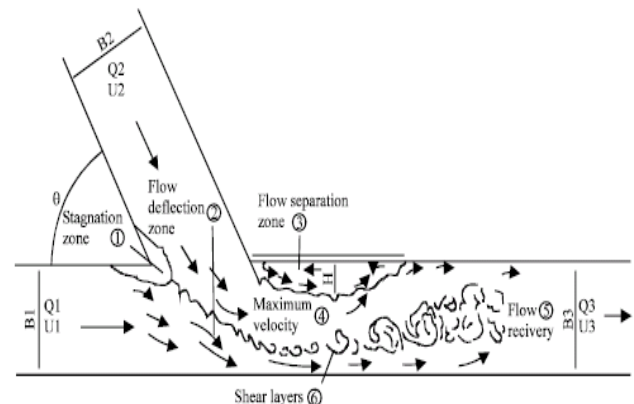

Figure 1: Observed zones by Best at the junction of two streams, [1].

A numerical modelling of shallow water in the presence of stationary and dynamic shocks was done by Ghezelsoflou and Jafarzadeh [2]. It is employed the second order Roe-TVD scheme for the numerical simulation of shallow flows both in one and two dimensions in the open channel. The numerical results are then compared with the experiments and the available analytical solutions. It may be deduced that, the scheme of Roe-TVD is a robust method for the simulation of complicated shocks in shallow water flows [3]. A comparison between the 1D and 2D numerical simulation of transitional flow in open-channel networks is presented and completely described allowing for a full comprehension of the modeling water flow. It is done by modelling flood in urban environment and finally, it obtained that second-order Runge–Kutta discontinuous Galerkin method is in a good conformity with experimental results. A modelling of water and sediment movement in dam break via finite volume method was applied by Baharestani and Banihashemi [4]. The assessment showed that it did not observe any numerical dispersion error adjacent to sharp gradient in all the tests which is one the capabilities of TVD in dam break modelling. Diaz et al studied the numerical approximation of bed load sediment transport due to water evolution [5]. For the hydro dynamical component, it was considered

Shallow Water equations. The morph dynamical component is defined by a continuity equation, which is defined in function of the solid transport discharge. Both components defined a coupled system of equations that could be rewrite as a non-conservative hyperbolic system. A two-dimensional scour model based on coupled system of shallow water equations (SWEs) and sediment transport on unstructured mesh is developed in order to solve coupled system of hydrodynamic and morph dynamic equations via finite volume method using Godunov scheme by Liu et al [6]. Two-dimensional (2D) bed-load transport simulations based on the depth-averaged shallow-water equations and the Exner equation were presented by Pacheco et al [7]. In this work, a Roe-type first-order upwind scheme has been applied as approximate Riemann solver for the discretization on 2D unstructured meshes. The models were tested comparing with exact solutions of 2D cases as well as experimental laboratory data. A study of a two-dimensional (2D) finite volume model that simulates sediment transport in overland and channel flow simultaneously was presented by Yu and Duan [8]. The model was based on the solutions of 2D shallow water equations coupled with the Exner equation. The 2D domain is discretized using Cartesian cells to formulate a Godunov-type cell-centered finite volume method. The numerical results showed that the model can satisfactorily reproduce the measured hydrographs and bed profiles for both dry bed and wet bed situations. The sediment transfer of rivers' bend and classification of the different bed particles was analyzed [9]. Flow pattern has been compared with numerical model in a constructed channel scaled 1:100. It indicates that average flow quantities are similar to numerical model but turbulent parameters differ from numerical model. Basri and Ghobadian presented the effect of downstream curved edge on local scouring at 60 degree open channel junction using SSIIM1 model. A 3D SSIM1 model was applied for assessing various channel discharge and width conditions. Eventually, the outcomes of experimental pilot was compared with numerical model. Flow pattern and sediment were considered at the rivers' interception by Habibi et al [11]. They evaluated the capability of the CCHE2D model base on available data in a laboratory. Comparing the lab and model outcomes demonstrated that the model can predict water profile at the interception of main and secondary branches on the condition that there is not any modelling of sediment where the model accuracy and the average of absolute error are 99 and 2.675 percent, respectively. Moreover, statistics indicated that the CCHE2D model is capable to simulate sediment longitudinal profile which accompanied by 99 percent precision and 99 percent coefficient of determination, as well as the average of absolute error was 15.2 percent.

Studies and researches are not sufficient in the flow pattern and sediment transfer fields where two channels encounter. In this paper, it is applied Roe finite volume method to assess the effect of the proportion of secondary channel width to main channel width on flow pattern and sediment transfer.

## II. SHALLOW WATER EQUATIONS

The shallow water equations obtain by assuming hydra static pressure distribution and also incompatible flow with averaging Navier-Stocks three dimensional in depth. The equations apply for studying a wide range of physical phenomenon such as dam break, flow in open channel, flood waves, forces on offshore structures and pollution transfer. The two dimensional form is [12, 13]:

$$\frac{\partial \xi}{\partial t}+\frac{\partial (uh)}{\partial x}+\frac{\partial (vh)}{\partial y} \quad (1)$$

$$\frac{\partial (uh)}{\partial t}+\frac{\partial (u^2 h)}{\partial x}+\frac{\partial (uvh)}{\partial y}-\nu\,(\frac{\partial (hu_x)}{\partial x}+$$
$$\frac{\partial (hu_y)}{\partial y}=\frac{\tau_{wx}-\tau_{bx}}{\rho}-gh\frac{\partial \xi}{\partial x}+hfv \quad (2)$$

$$\frac{\partial (uh)}{\partial t}+\frac{\partial (uvh)}{\partial x}+\frac{\partial (v^2 h)}{\partial y}-\nu\,(\frac{\partial (hv_x)}{\partial x}+$$
$$\frac{\partial (hv_y)}{\partial y}=\frac{\tau_{wy}-\tau_{by}}{\rho}gh\frac{\partial \xi}{\partial y}+hfu \quad (2)$$

Where $\xi$ is the height above the water level which is $h_s$, total depth become $h_s+\xi$, the u and v are the average velocities along the x and y axes, the t is time, the $\tau_{wx}$ and $\tau_{wy}$ are wind shear tensions, the $\tau_{bx}$ and $\tau_{by}$ are bottom friction forces, the $\nu$ is viscosity, the g is gravity and the *f* is Coriolis parameter, respectively.

## III. NUMERICAL SOLUTION BY FINITE VOLUME METHOD

Shallow water equations are used in complicated situations such as rivers. Regarding, adopting the unstructured components are appropriate because they can coordinate better with boundaries. In this study, by discretizing equations into time and using a semi step, finally, it is applied second-order method into time. Furthermore, it uses triangle cells and Roe method to discretize the equations locally. In order to achieve second order precision, it is employed multi-dimensional slope restriction, locally [14-16]. The Grass equation is used to simulate sediment transfer:

$$q_b = A_g \cdot \frac{q_f}{h} \left\| \frac{q_f}{h} \right\|^{m_g - 1} \quad (4)$$

Where $q_b$ is sediment discharge, $A_g$ and $m_g$ are in the rage of $1 \geq m_g < 4$ and $0 > A_g > 1$ base on flow and sediment properties. $A_g$ coefficient obtains from experimental data. It depends on particles size and flow viscosity and the weaker fluid and sediment interaction is, the value becomes smaller.

## IV. EVALUATION OF NUMERICAL MODEL

In order to evaluate the numerical code, it is applied Ghobadian model at junction of two channels which the degree junction and slope are 60o and 0, respectively [10]. The main and secondary channels length considered 9 and 3 m and also their widths are 35 and 25 cm, respectively. The inlet discharge to the pilot is measured by an electronic flow meter which its precision is 0.01liter per sec. Applied sediment would be constant classification with particles' average diameter of 1.95 mm and GS=2.65. At first, 11.5 cm of sediment materials are placed at the bottom of v and secondary flume. Fig, 2 shows a layout of applied equipment in laboratory.

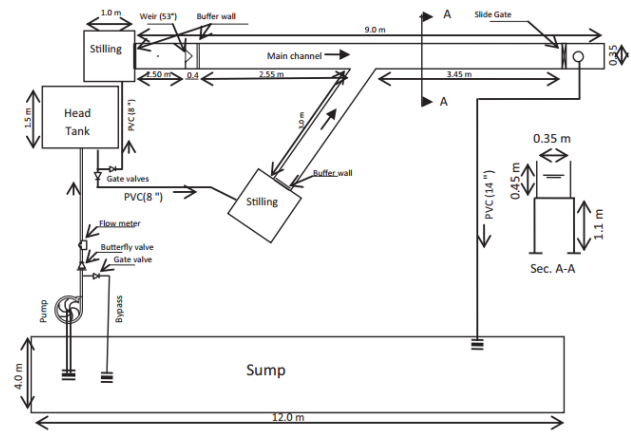

Figure 2: Layout of applied equipment in laboratory

In the numerical model, defined conditions were applied and then assigned 5000 numbers of mesh to obtain optimum cells number. Courant-Friedrichs-Lewy number and water upstream depth were considered 0.9 and 15 cm, respectively. The Manning coefficient is applied 0.009. The numerical model has done with regard the proportion of main to secondary channel discharge to 0.66. It is defined sediment properties to simulate sediment behavior. In this regard, it is considered ultimate shear tension to 0.047 according to the diameter of sediment particles. It is observed that numerical model becomes convergent after lasting 220 sec by applying required value and coefficient for sediment models and comparing obtained outcomes from sediment and flow assessment. Fig, 1 presents the results of bed variations in numerical model during 220 sec and experimental model. Fig, 2 shows the effect of discharge ratio to final height of sedimentation in both numerical and experimental models.

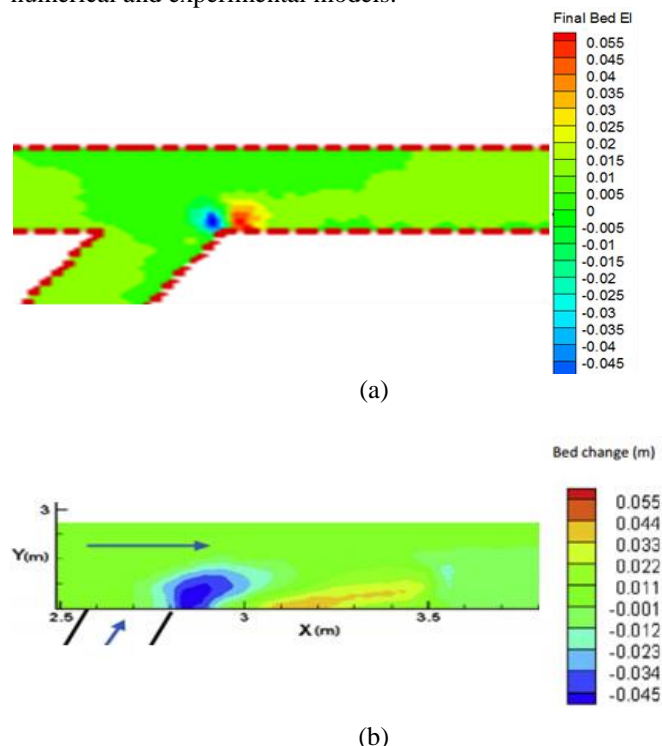

(a)

(b)

Figure 3: Bed variations results, a) numerical model b) experimental

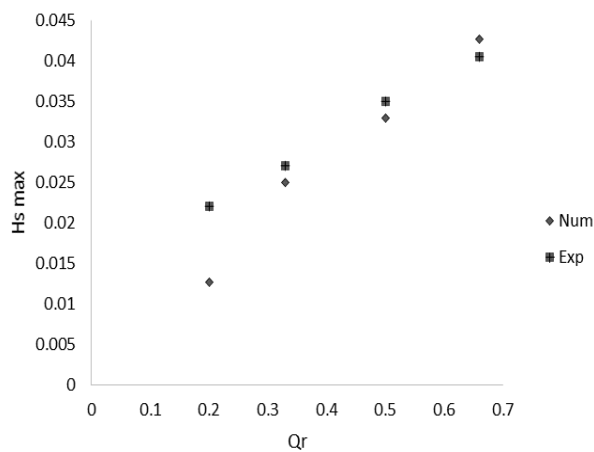

Figure 1: The effect of discharge ratio to final height of sedimentation in both numerical and experimental models

TABLE I
PRESENTS FINAL VALUES OF SEDIMENTATION HEIGHT IN BOTH NUMERICAL AND EXPERIMENTAL MODELS IN DIFFERENT DISCHARGE RATIO AND ABSOLUTE VALUE OF RELATIVE ERROR

| Relative error (%) | $Hs_{max}$ num (m) | $Hs_{max}$ exp (m) | Qr |
|---|---|---|---|
| 74.6 | 0.0126 | 0.022 | 0.2 |
| 8 | 0.025 | 0.027 | 0.33 |
| 6.06 | 0.033 | 0.035 | 0.5 |
| 4.9 | 0.0427 | 0.0406 | 0.66 |

According to Table, 1 and Fig, 4, maximum bed variation of numerical model with a minor difference is convergent with experimental model and the results are then acceptable.

## V. ASSESSING THE EFFECT OF HYDRAULIC CONDITION ON CROSS CHANNEL

The geometry model which used, is including of two cross channels where they intercept each other by 60 degree angle and also bottom slope is equal to zero. The main and secondary channels length considered 9 and 3 m and also their widths are 35 and 25 cm, respectively. Applied sediment would be constant classification with particles' average diameter of 1.95 mm and $G_S$=2.65 which they cover 11.5 cm depth of the flume. According to assessments, solution space is divided by 5400 cells. Flow height is defined 15 cm in both main and secondary channels and Manning coefficient is set to 0.009. Fig, 5 presents the layout of studied model.

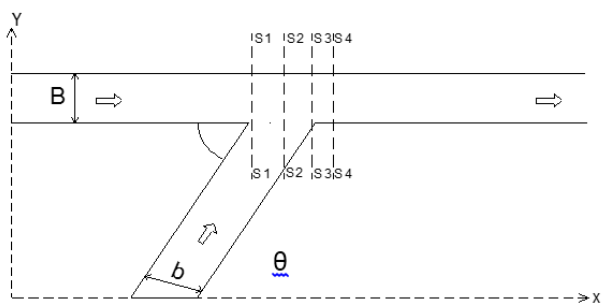

Figure 2: Layout of studied model

*On Flow Pattern and Sediment Transfer*
One the most effective and important parameters in flow pattern, erosion and sedimentation is secondary discharge channel to total discharge ratio. In this paper, the effect of four ratios like 0.2, 0.33, 0.5 and 0.66 has been studied.

*Flow Pattern*
Flow velocity distribution diagram accompanied by flow lines in different discharge ratios observe along x axis in Fig, 6(a to d).

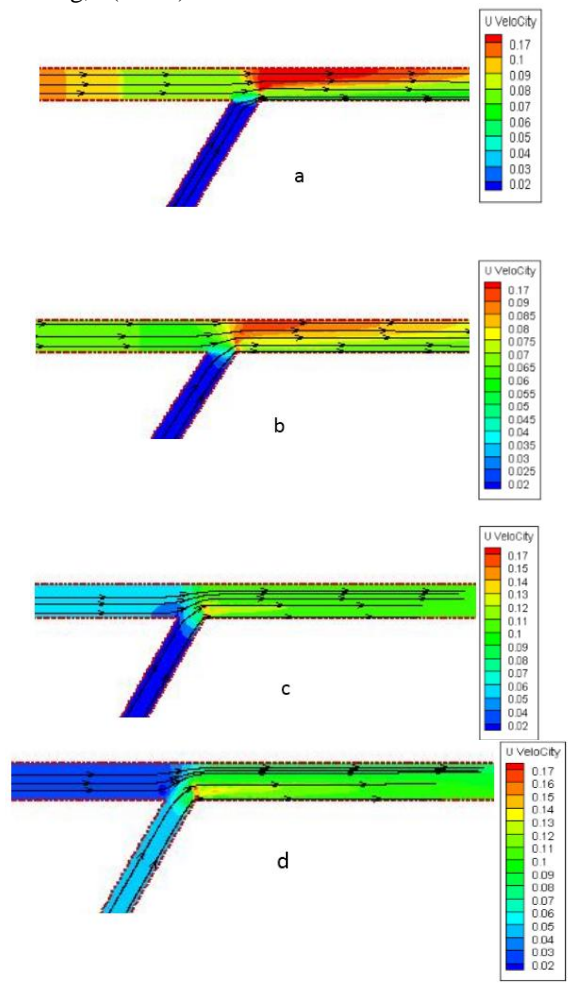

Figure 3: Velocity diagram in x direction with discharge ratios, a) 0.2 b) 0.33 c) 0.5 d) 0.66

It is observed that velocity in secondary channel and at the beginning of main channel is less in all the shapes and it is then indicated that maximum velocity has occurred at the junction. Study reveals that velocity is bigger than the rest of discharge ratios as it is 0.2. In this case, maximum velocity happens in the main channel at the junction. Velocity decreases as discharge ratio is increasing along the channel. In all the shapes, velocity rises at the junction.

*Bed Longitudinal Profile*
The results of bed variations has been shown in different discharge ratios along the channel in Fig, 7. Also, maximum bed height observes in different discharge ratios in Fig, 8.

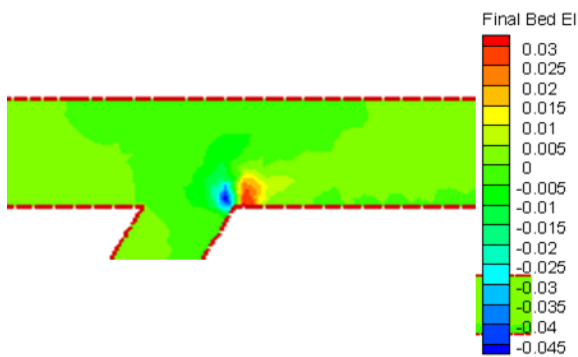

Figure 4: Bed variation plan in 0.5 discharge ratio

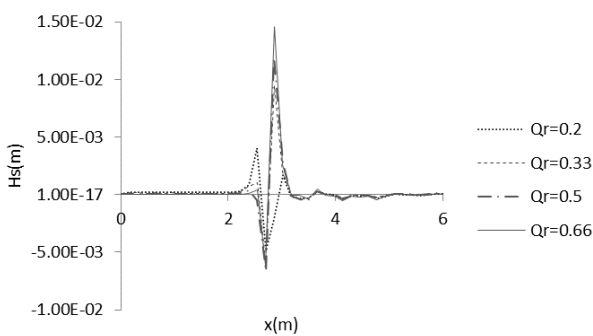

Figure 5: Variations of channel longitudinal profile in y=2.75 faced with final bed in different discharge ratios

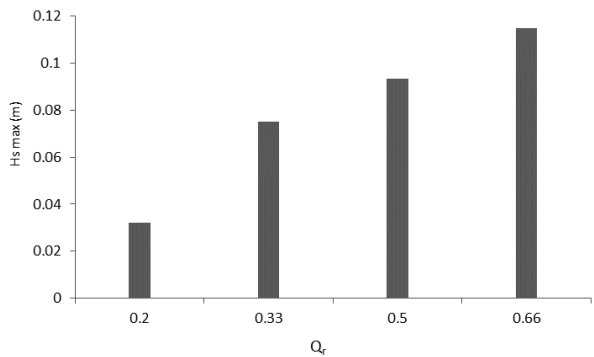

Figure 6: Pie chart of maximum bed height in different discharge ratios

It presents that maximum bed height occurs in 0.5 discharge ratio at the junction which appears like erosion and sedimentation (Fig, 7). It shows that bed variations increase as secondary channel discharge to total discharge ratio is rising. Increasing the depth of erosion can be caused by increasing momentum secondary branches and ultimately, speed up the inlet flow from secondary channel to the junction which causes increasing shear tension and the intensity of formed vortices at the bottom of corner of junction. Furthermore, one of the reasons of increment in sedimentation height could be the impact of increasing in separation to sedimentation zone dimensions. Fig, 9 also shows that increment in maximum discharge ratio accompanies by rising bed height. The results reveal that bed height has increased 32 percent by rising 51 percent in maximum discharge, on average.

***Evaluating of Discharge Ratio on Bed Cross Profile***

The outcomes of bed variations through the width channel in s4-s4 and s3-s3 cross sections has been shown for different discharge ratios in Fig, 10 (a and b).

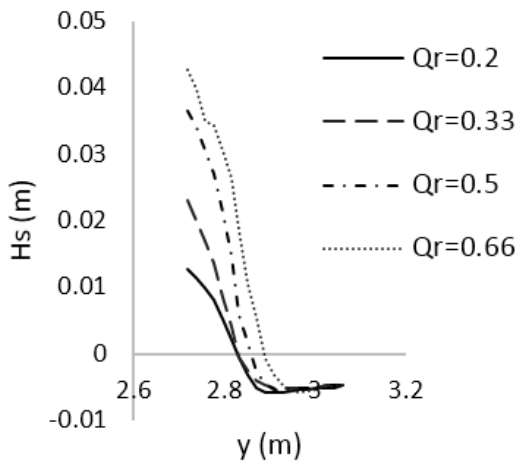

(a)

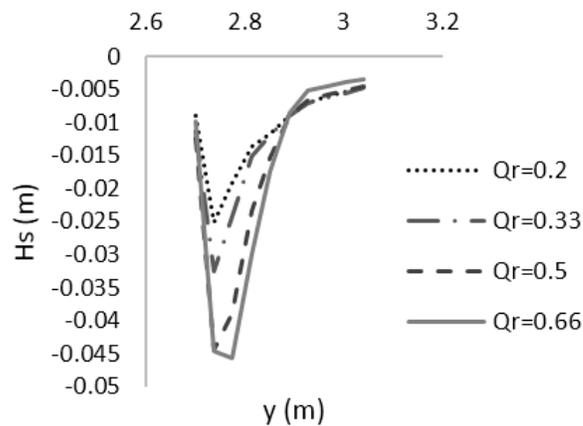

(b)

Figure 7: Channel cross profile variations faced with bed height in discharge ratios

It is observed that variations of final bed channel increases by increment in discharge ratio around the junction in a case and variations of discharge ratio does not effect on bed variations above the secondary channel. The b case shows erosion depth goes up as discharge ratio rises. Minimum point of discharge ratios has occurred in a certain distance to channel entrance which bed height

                      

variations starts from -0.01 and the point of curves interception is x=3m in all the studied cases.

***Evaluating of Froude Number on Bed Height***

Another parameter, which is influential on sediment pattern, is Froude number. Fig, 11 presents the results of Froude number of secondary to main channel ratio on the maximum height of sedimentation and sediment-transfer.

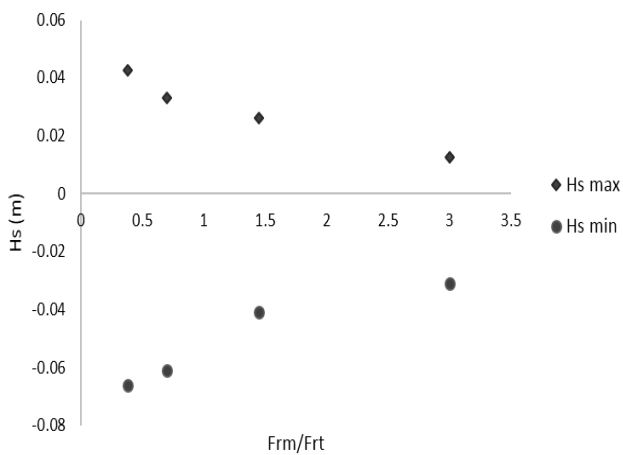

Figure 8: Effect of Froude number on the height of sedimentation and sediment-transfer

Where $F_{rm}$ and $F_{rt}$ are Froude numbers at the beginning of main and secondary channels, respectively. The results show that maximum height of sedimentation creases by declining in the Froude number ratio. Moreover, increasing the ratio accompanies by decreasing bed erosion value.

## VI. EVALUATING THE EFFECT OF GEOMETRY

***Flow Pattern and Sediment Transfer***

One of the most effective and important parameters on the height of sedimentation and erosion is secondary to main channel width ratio and evaluating the effect of secondary to main channel width ratio on flow pattern and sediment transfer. In this step, three ratios such as 0.42, 0.71 and 1 in the above channel is assessed.

***Velocity Flow Ratio***

The results of secondary to main channels widths ratio on flow velocity has been demonstrated in 2.7m distance to the secondary channel entrance in cross direction along the channel in Fig, 12. Where b and B are secondary and main channel widths, respectively.

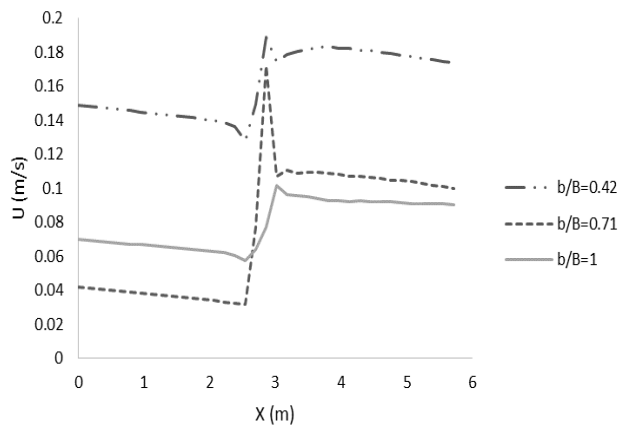

Figure 9: Effect of secondary to main channel width ratio on flow velocity along the channel

According to Fig, 12, it is faced with declining in velocity on three width ratios before secondary channel and it is observed increment in velocity at the junction and ultimately, velocity decreases as it has passed the junction. It should be considered that flow velocity after the junction is bigger than the junction in all the cases. Velocity variation trend is more in the 0.71 ratio rather than the two other ratios, on comparison. Moreover, flow velocity decreases as the ratio increases from 0.42 to 1.

***To Final Bed Height Ratio***

It has presented the effect of width to final bed height ratio along the channel in Fig, 13

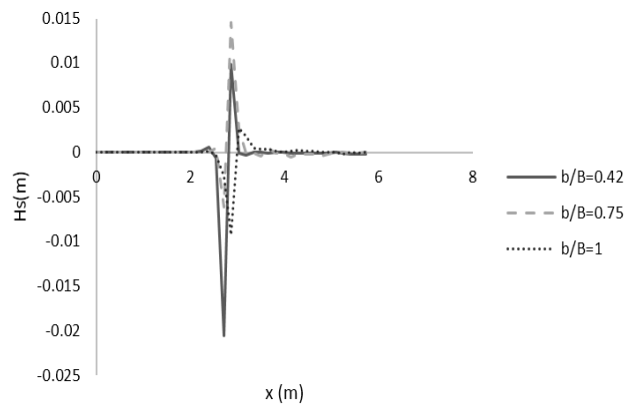

Figure 10: Effect of secondary to main channel width ratio on final bed height along the channel in y=2.7m.

Bed height is constant before and after the junction in the main channel and bed variations is similar in all three ratios. Decreasing and increasing in bed height occur at the entrance and exit channel and also Maximum values of erosion and sedimentation has occurred in 0.42 and 0.75 ratios, respectively. Fig, 14 shows bed height variations in cross section of the channel faced with different width ratios in s2-s2 cross section obtained from Fig, 5. Bed height variations in the cross section of the channel also is presented in Fig, 15 in s3-s3 cross section obtained from Fig, 5.

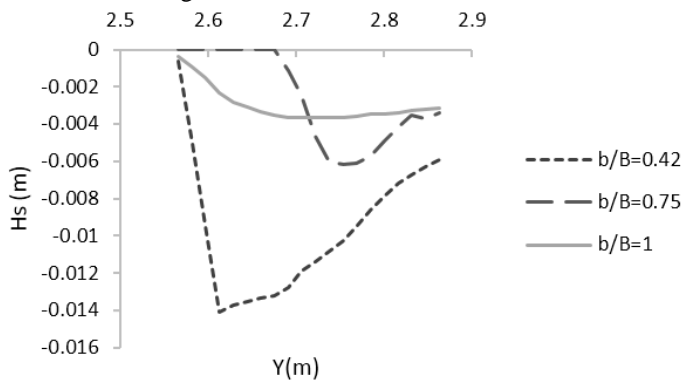

Figure 11: Bed height variations in width ratios in s2-s2 cross section

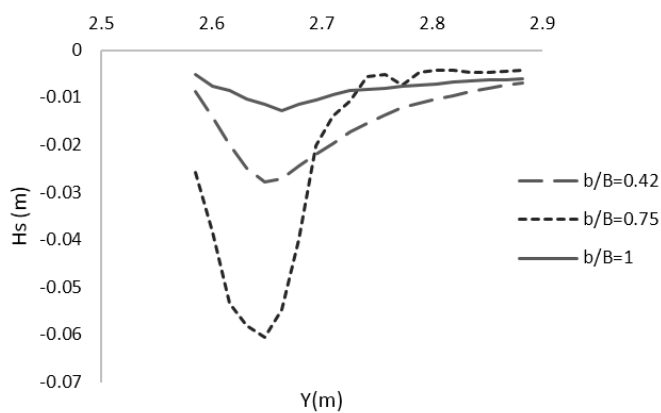

Figure 12: Bed height variations in width ratios in s3-s3 cross section

Fig, 14 indicates that at first, bed height decreases and then increases in both ratios of 0.42 and 1 in the s2-s2 cross section. Increment of bed height has happened in the mid position of main channel. In the ratio of 0.75, bed variations is marginal and afterwards falling. Ultimately, the bed height is increasing. In all the width ratios, it falls and then rises in Fig, 15. The bed height variations is bigger in the 0.75 ratio rather than the rest two ratios. Whatever it is gone away from secondary channel, decreased trend becomes less.

### *Effect of Junction Angle*
One of the most important parameter on flow pattern and sedimentation is junction angle in cross channels. In this paper, it is studied three angles such as 60o, 75o and 900.

### *Velocity Pattern*
Fig, 16 from a to c shows flow lines in three angles of 60o, 75o and 900 during 50 sec, respectively.

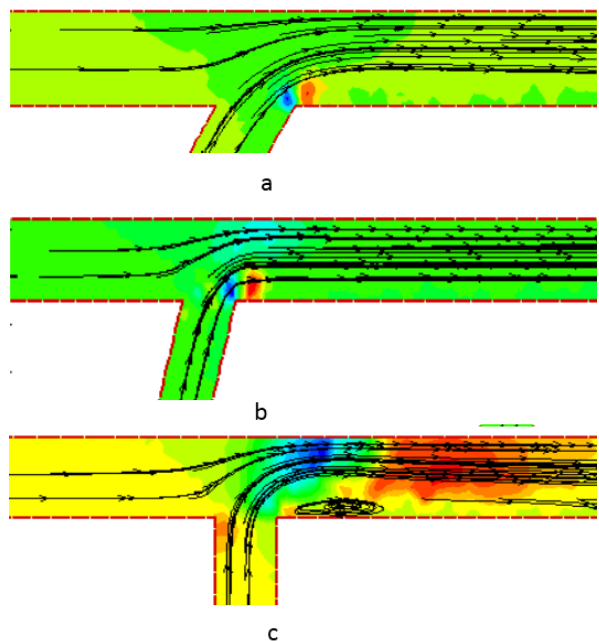

Figure 13: Flow lines in three angles a) 60 b) 75 and c) 90

According to above figures, rotating flow occurs after channel junction in 900 due to increment in flow separation zone and junction angle as well as creation of bigger vortices. It is presented flow velocity variations in different angles along the channel in Fig, 17.

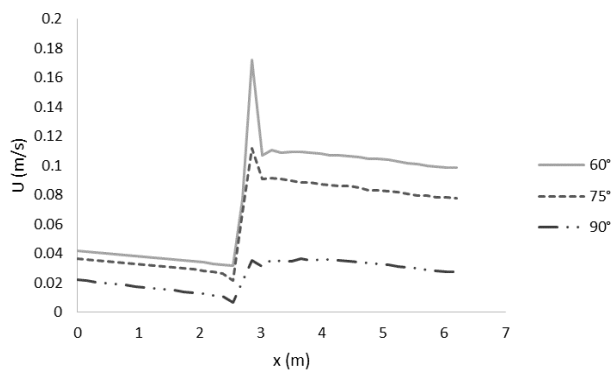

Figure 17: Effect of angle on flow velocity along the channel

It is observed that the more angle increases, the more velocity decreases along the channel. Maximum variations happens at the beginning of the junction to last it. It is faced with declining in velocity and then rises at the beginning of the junction and ultimately, velocity decreases to the last of the main channel.

*Erosion and Sedimentation*

Varying the junction angle effects on the length of erosion and sedimentation, as well. Fig, 18 depicts studied results of effect of junction angle on bed depth along the channel.

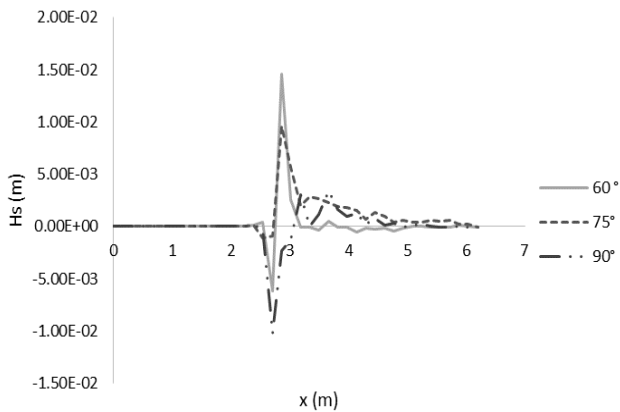

Figure 14: Longitudinal profile of junction angle influence on bed depth

Bed depth increases as junction angle is rising as well as flow entrance angle ascends. It is caused flow of secondary channel keeps out from main channel wall. Consequently, separation zone enlarges and the more intensive vortices create. Provided that the rotation intensity of vortex flow becomes more, erosion increases. Bed variations has been shown in Fig, 19 and 20 in s1-s1 and s3-s3 cross sections obtained from Fig, 5.

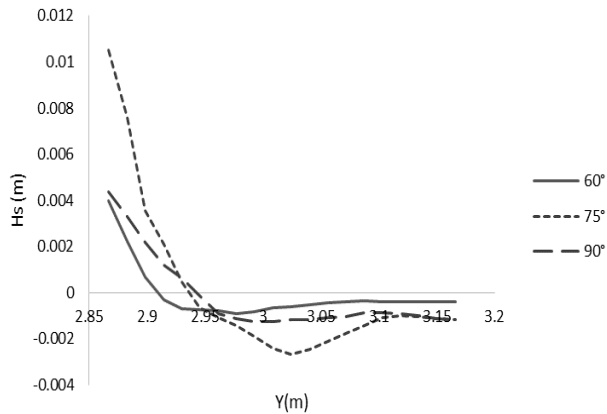

Figure 19: Bed variations height in cross section (s1-s1)

The results of above figure shows that at first, bed height decreases and then rises in given cross section in all three angles. Also, sedimentation and erosion happen in secondary and main channel, respectively. Moreover, bed height variation is more in $75^o$ rather than the rest two angles.

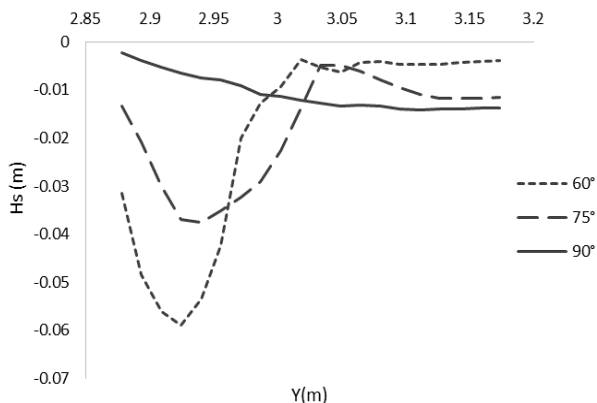

Figure 20: Bed variations height in cross section (s3-s3)

According to above figure, bed erosion happens in three angles. It is demonstrated that bed height falls and then rises in $60^o$ and $75^o$, respectively. Eventually, it decreases. It is observed declining in bed height in $90^0$ as well as variation trend is constant. The less and the more bed variation belongs to $90^0$ and $60^0$, respectively.

## VII. CONCLUSION

In this paper, it is used a written code based on Roe-finite volume method to study the effect of secondary to main channel width on flow, erosion and sedimentation. Considerable results obtained by comparing the experimental data with model outcomes. Regarding, numerical modelling has been set in cross channel. The results of discharge ratio reveals that as it is increasing, bed variation height develops. Cross section variation after secondary channel shows that increment in discharge ratio accompanies with ultimate bed variation near the junction. Assessment of secondary to main channel Froude number indicates that maximum height of sedimentation decrease by declining in the correspond ratio. Ultimate bed height and velocity variations change by given trend in three ratios of secondary to main channel width along the channel. On the evaluating of junction angle, it shows that increment in angle has converse relationship with flow velocity variations. Furthermore, velocity and topographical bed variations follow a specific pattern in all the studied angles.